\documentclass[twocolumn,secnumarabic,amssymb, amsmath, nofootinbib,tightenlines,
nobibnotes, aps, prl,epsfig]{revtex4}
\usepackage{graphicx}
\usepackage{dcolumn}
\usepackage{bm}
\begin{document}
\preprint{APS/123-QED}
\title{Analysis of the proton longitudinal structure function from the gluon distribution function }

\author{G.R.Boroun}%
 \email{grboroun@gmail.com; boroun@razi.ac.ir }
\author{B.Rezaei }
\altaffiliation{brezaei@razi.ac.ir}
\affiliation{ Physics Department, Razi University, Kermanshah
67149, Iran}
\date{\today}
\begin{abstract}
We make a critical, next- to- leading order, study of the
relationship between the longitudinal structure function $F_{L}$
and the gluon distribution proposed in Ref.[2], which is
frequently used to extract the gluon distribution from the proton
longitudinal structure function at small $x$. The gluon density is
obtained by expanding at particular choices of the point of
expansion and compared with the hard pomeron behaviour for the
gluon density. Comparisons with H1 data are made and predictions
for the proposed best
approach are also provided.\\
\end{abstract}
 \pacs{***}
\maketitle
\subsection{1.Introduction}
The longitudinal proton structure function $F_{L}$, measured in
the deep inelastic lepton- proton scattering, is proportional to
the cross section for the interaction of the longitudinally
polarized virtual photon with a proton. This observable is of
particular interest since it is directly sensitive to the gluon
density. In the quark- parton model, the longitudinal structure
function is zero since longitudinal polarized photons do not
couple to spin $\frac{1}{2}$ quarks. In the lowest order DGLAP [1]
approximation of QCD, the longitudinal structure function is given
by
\begin{equation}
F_{L}(x,Q^{2})=K^{S+NS}{\otimes}F_{2}(x,Q^{2})+K^{G}{\otimes}G(x,Q^{2}),
\end{equation}
where both quarks and gluons contribute. The variable $Q^{2}$ is
the square of the four- vector momentum exchange, and $x$ is the
Bjorken scaling variable. Here kernels ($K^{i}$) are the
coefficients functions and the symbol ${\otimes}$ denotes
convolution according to the usual prescription. At small values
of $x$, $F_{L}$ is driven mainly  by gluons through the transition
$g{\rightarrow}{\hspace{0.1cm}}q\overline{q}$ ($g(x,Q^{2})$ is the
gluon density). Therefore $F_{L}$ can be used for the extraction
of the gluon distribution in the proton, and it provides a crucial
test of the validity of perturbative QCD in this kinematical
range. The knowledge of $F_{L}$ gives a complementary
determination of the gluon distribution to that using the scaling
violation of $F_{2}$, since at low $x$ we have
\begin{equation}
F_{L}^{g}(x,Q^{2})=K^{G}_{LO+NLO}{\otimes}G(x,Q^{2}).
\end{equation}
In this paper we deduce the general relations between the
longitudinal structure function and the gluon distribution
function with analytical methods at NLO. In Ref.[2] a relationship
between $F_{L}$ and the gluon was proposed in order to facilitate
the extraction of the gluon density from data. We demonstrate here
that the validity of this relation crucially depends on the choice
of expansion point. Our purpose here is to improve the situation
with a new expansion of the gluon distribution function at LO and
NLO. In addition our results provide a useful NLO expression valid
at small $x$, where the gluon has a hard- Pomeron behaviour.
Section 2 is devoted to a revision of the method proposed in
Ref.[2] and gives a general expression at NLO. Finally, in section
3, we present our
conclusions.\\
\subsection{2.Compact Formula}
The standard collinear  factorization formula for the longitudinal
structure function at low $x$ reads
\begin{equation}
F_{L}^{g}(x,Q^{2})=\int_{0}^{1-x}\frac{dz}{1-z}K^{G}_{LO+NLO}(1-z)G(\frac{x}{1-z},Q^{2}),
\end{equation}
where $F_{L}$ is derived from the integrated gluon distribution.
The analytical expression of the gluon kernel [3-4] at LO and NLO
is defined in the Appendix. An approximate relationship between
$F_{L}$ and the gluon distribution can be derived from the
expansion of $G(y,Q^{2})$ around a chosen expansion point.
Therefore, the choice of the
 point of expansion is associated with the relation between of
 $x_{g}$ and $x$.\\

$\bf{A}$): In Ref.2 the authors have suggested that expression (3)
at LO can
 be reasonably approximated by
\begin{equation}
F_{L}^{g}(x,Q^{2})=\frac{2\alpha_{s}}{\pi}\frac{\sum_{i=1}^{N_{f}}e_{i}^{2}}{5.9}G(2.5x,Q^{2})
\end{equation}
which demonstrates the close relation between the longitudinal
structure function and the gluon distribution at small $x$.\\

$\bf{B}$): Using the expansion method for the gluon distribution
function at an arbitrary point $z=a$ as
\begin{equation}
\frac{x}{1-z}|_{z=a}=\frac{x}{1-a}\sum_{k=1}^{\infty}[1+\frac{(z-a)^{k}}{(1-a)^{k}}].
\end{equation}
The above series is convergent for $|z-a|<1$. According to this
expression we can expand the gluon distribution $G(\frac{x}{1-z})$
as
\begin{eqnarray}
G(\frac{x}{1-z})|_{z=a}&=&G(\frac{x}{1-a})\\\nonumber
&&+\frac{x}{1-a}(z-a)\frac{{\partial}G(\frac{x}{1-a})}{{\partial}x}+O(z-a)^{2}.
\end{eqnarray}
Retaining terms only up to the first derivative in the expansion
and doing the integration, we obtain our master formula as
\begin{eqnarray}
F_{L}^{g}(x,Q^{2})=A(x,Q^{2})
{\times}G(\frac{x}{1-a}(1-a+\frac{B(x,Q^{2})}{A(x,Q^{2})})),
\end{eqnarray}
where
\begin{eqnarray}
A(x,Q^{2})=\int_{0}^{1-x}\frac{1}{1-z}K^{G}_{LO+NLO}(1-z)dz,
\end{eqnarray}
and
\begin{eqnarray}
B(x,Q^{2})=\int_{0}^{1-x}\frac{z-a}{1-z}K^{G}_{LO+NLO}(1-z)dz,
\end{eqnarray}
where  $\alpha$ has an arbitrary value $0{\leq}\alpha{<}1$.\\

Eq.7 can be rewritten as
\begin{eqnarray}
F_{L}^{g}(x,Q^{2})=\frac{10\alpha_{s}}{27\pi}G(\frac{x}{1-a}(\frac{3}{2}-a)),
\end{eqnarray}
In the limit $x{\rightarrow}0$ for LO analysis. This result shows
that the longitudinal structure function $F_{L}^{g}(x,Q^{2})$ at
$x$ is calculated using the gluon distribution
$\frac{x}{1-a}(\frac{3}{2}-a)$ at the limit $x{\rightarrow}0$.
When the points $a=0$, $a=0.666$ and  $a=0.9$ are used, we get
respectively
\begin{eqnarray}
F_{L}^{g}(x,Q^{2})=\frac{10\alpha_{s}}{27\pi}G(\frac{3}{2}x)(<Ref.[2]),
\end{eqnarray}
\begin{eqnarray}
F_{L}^{g}(x,Q^{2})=\frac{10\alpha_{s}}{27\pi}G(2.5x)(=Ref.[2]),
\end{eqnarray}
and
\begin{eqnarray}
F_{L}^{g}(x,Q^{2})=\frac{10\alpha_{s}}{27\pi}G(6x)(>Ref.[2]).
\end{eqnarray}

Therefore these equations show that our approximation differs from
the result of Ref.[2] . In the expanding the gluon distribution
around $z=0$, the result (11) differs from Ref.[2] result. We
explain that difference arises in the expansion method. Our
results cover expanding of the gluon distribution at all points of
the
expansion.\\
In Fig.1  we present the results of the longitudinal structure
function at some points of expansion $z=a$ using the gluon
distribution, which is usually taken from the  Block [5] model at
LO. The DL [6-7] fit and GRV [8] parameterizations have also been
used to investigate quantitatively the accuracy of $F_{L}$ with
respect to the gluon distribution. We introduced  the fractional
accuracy variable
\begin{equation}
 Fractional~ Accuracy ~{\equiv}~
F_{i,MRST}~  {\equiv}~ 1-\frac{F_{L}(i)}{F_{L}(MRST)}
\end{equation}
 where $i=GRV, DL$ and
$Block$ distributions with $F_{L}(MRST)$ denoting the predictions
of the longitudinal structure function in NLO PQCD which are given
by the MRST parametrizations [11]. We show in Fig.2 that the
fractional accuracy for the the Block model is the best which
confirms the correctness of our solution for $F_{L}$ with Block
model. We also note that there is disagreement between Block
distribution and NLO-QCD fit to H1 data. It is clear that Block
starting distribution is at LO when compared with H1 data in
an NLO analysis.\\
 Therefore, in Fig.1, we compare our results at
two points of the expansion $a=0$ and $a=0.9$ with the the gluon
distribution input according to the Block model [5]. We compared
our results with the results of Ref.[2], H1 data [9] and MVV
prediction [10]. These results show that the expansion of the
gluon distribution around $a=0$ differs from the Ref.[2] result,
and is equal to that result only at the point of expansion
$a=0.666$. But the best results can be found in the range of
expansion at the points of $0.7{\leq}a{\leq}0.95$. It is easy to
see that the validity of the expansion in LO depends strongly on
the point of the expansion of the gluon distribution and this is
comparable with the experimental data only at high
values of the point of expansion.\\

We also obtain the relation between the longitudinal structure
function and gluon distribution exact to  NLO analysis which
adopts the similar integral form (Eqs.7-10), where the coefficient
function $K^{G}$ is well known up to NLO. Therefore, an explicit
relation between the functions can be found at this order.
Unfortunately there is not an analytical form at this order.
Consequently, in order to estimate the magnitude of the NLO
corrections for the low $x$ range, we have numerically computed
$F_{L}^{g}(x,Q^{2})$ at a moderate $Q^{2}=20 GeV^{2}$.\\
In order to show this we firstly computed the relation between
$x_{g}$($=kx$, $k=\frac{1}{1-a}(1-a+\frac{B}{A})$) and $x$ for
three different of $x$ values at $Q^{2}=20 GeV^{2}$ in Fig.3.
These distributions represent the whole spectrum of possible
behaviours of $x_{g}$ with respect to the points of the expansion
at small $x$. They tend to constant values at small $x$ at NLO .
Having done that we determined the longitudinal structure function
at point of expansion $a=0$ and the point of expansion which
corresponds to the maximal deviation from the result for $a=0$ for
several
chosen $x$ values. Since these latter points of expansion differ they are separately labelled on the Fig.4.\\
The results of this analysis are shown in Fig.4. We plot there the
lower and higher limit of the point of expansion at $Q^{2}=20
 GeV^{2}$ in NLO analysis. It is easy to see that the validity of
relation (7) in NLO depends strongly on the lower limit of the
point of expansion. By contrast for the NLO analysis, the relation
is still approximately true for the points of expansion bigger
than $0.7$. We can conclude that, when compared with the behaviour
at LO, we cannot find a monotonic relation between $x_{g}$ and $x$
at NLO but comparison to H1 data suggests that the expansion
point $a=0$ yields the best results.\\

$\bf{C}$): Exploiting the low- $x$ behaviour of the gluon
distribution function according to the hard (Lipatov) Pomeron as
\begin{equation}
G(x,Q^{2})= f(Q^{2})x^{-\delta}.
\end{equation}
The power of $\delta$ is found to be either $\delta {\simeq} 0$ or
$\delta{\simeq} 0.5$. The first value corresponds to the soft
Pomeron and the second value to the hard (Lipatov) Pomeron
intercept. Based on the hard (Lipatov) pomeron behaviour for the
gluon distribution, let us put Eq.(15) in Eq.(3). After doing the
integration over $z$, Eq.(3) can be rewritten as
\begin{eqnarray}
F_{L}^{G}(x,Q^{2})=I_{G}(x,Q^{2}){\times}G(x,Q^{2}),
\end{eqnarray}
where
\begin{eqnarray}
I_{G}(x,Q^{2})=\int_{0}^{1-x}k_{LO+NLO}^{G}
(1-z)(1-z)^{\delta-1}dz.
\end{eqnarray}
 We observe that equation 16  implies a relationship between $F_{L}$ and the gluon
 at the same value of $x$.\\
In Fig.5 we present results for $F_{L}$ at LO and NLO using the
gluon distribution [5]. This seems to indicate that $F_{L}$ is
dominated
 at small $x$ by  hard Pomeron exchange. This powerful approach to the
 small-$x$ data for $F^{g}_{L}(x,Q^{2})$  extends the Regge
 phenomenology that is so successful for hadronic processes. The
 Ref.[2] result (solid line) is also shown. We immediately see
 that our results at LO and NLO with respect to the hard Pomeron
 behaviour have much better  behaviour than the Ref.[2] result.\\

 \subsection{3.Conclusions}
In this work we discuss the approximative determination of the
longitudinal structure function at low $x$ with respect to the
expansion of the gluon distribution at arbitrary point of
expansion of $G(\frac{x}{1-z})$ at LO and NLO. When comparing
results, we conclude that the more suitable points of expansion
are in the range $0.7{\leq}a{\leq}0.95$ at LO and $a{\simeq}0$ at
 NLO. We demonstrate that  the LO relation (4)
between the longitudinal
 structure function and the gluon distribution is not generally
 valid. It remains a reasonable
 approximation only at the point of expansion $a=0.666$ for a gluon
 distribution which is sufficiently singular. At NLO the H1 data indicate that the optimal point of expansion
is  $a{\simeq}0$. Fortunately, the hard Pomeron behaviour for the
gluon distribution
 estimates $F_{L}$ at LO and NLO with
 reasonable validity for $x_{g}=x$.\\

$\bf{Acknowledgements}$   G.R.Boroun would like to thank the
anonymous referee of the paper for his/her careful reading of the
manuscript and for the productive discussions.\\

\newpage
\subsection{Appendix}
The explicit form of the gluon kernel is given by the following:
\begin{widetext}
\begin{eqnarray}
K_{LO}^{G}(\frac{x}{y},Q^{2})&=&\frac{\alpha_{s}}{4\pi}[8(x/y)^{2}(1-x/y)][\sum_{i=1}^{N_{f}}e_{i}^{2}]\nonumber\\
K_{NLO}^{G}(\frac{x}{y},Q^{2})&=&(\frac{\alpha_{s}}{4\pi})^{2}[\sum_{i=1}^{N_{f}}e_{i}^{2}][16C_{A}(x/y)^2(+4dilog(1-x/y)\hspace{1cm}\nonumber\\
&&-2(1-x/y)ln(x/y)ln(1-x/y)+2(1+x/y)dilog(1+x/y)+3ln(x/y)^2\nonumber\\
&&+2(x/y-2)Pi^2/6+(1-x/y)ln(1-x/y)^2+2(1+x/y)ln(x/y)ln(1+x/y)\nonumber\\
&&+\frac{(24+192x/y-317(x/y)^2)}{24(x/y)}ln(x/y)+\frac{(1-3x/y-27(x/y)^2+29(x/y)^3)}{3(x/y)^2}ln(1-x/y)\nonumber\\
&&+\frac{(-8+24x/y+510(x/y)^2-517(x/y)^3)}{72(x/y)^2})-16C_{F}(x/y)^2(\frac{5+12(x/y)^2}{30}ln(x/y)^2\nonumber\\
&&-(1-x/y)ln(1-x/y)+\frac{(-2+10(x/y)^3-12(x/y)^5)}{15(x/y)^3}(+dilog(1+x/y)\nonumber\\
&&+ln(x/y)ln(1+x/y))+2\frac{5-6(x/y)^2}{15}Pi^2/6+\frac{4-2x/y-27(x/y)^2-6(x/y)^3}{30(x/y)^2}ln(x/y)\nonumber\\
&&+\frac{(1-x/y)(-4-18x/y+105(x/y)^2)}{30(x/y)^2})].
\end{eqnarray}
\end{widetext}
For the SU(N) gauge group, we have $C_{A}=N$,
$C_{F}=(N^{2}-1)/2N$,
 $T_{F}=N_{f}T_{R}$, and $T_{R}=1/2$ where $C_{F}$ and $C_{A}$ are the color Cassimir
 operators.
\section{References}

1.Yu.L.Dokshitzer, Sov.Phys.JETP {\textbf{46}}, 641(1977);
G.Altarelli and G.Parisi, Nucl.Phys.B \textbf{126}, 298(1977);
V.N.Gribov and L.N.Lipatov, Sov.J.Nucl.Phys. \textbf{15},
438(1972).\\
2.A.M.Cooper-Sarkar et.al., Z.Phys.C{\bf39}, 281(1988);
Acta.Phys.Polon.B{\bf34}, 2911(2003).\\
3. J.L.Miramontes, J.sanchez Guillen and E.Zas, Phys.Rev.D \textbf{35}, 863(1987).\\
4. D.I.Kazakov, et.al., Phys.Rev.Lett. \textbf{65}, 1535(1990).\\
5.M.M.Block, et.al., Phys.Rev.D{\bf77},094003(2008).\\
6. A.Donnachie and P.V.Landshoff, Phys.Lett.B {\bf437}, 408(1998 ).\\
7. A.Donnachie and P.V.Landshoff, Phys.Lett.B {\bf550}, 160(2002 )\\; P.V.Landshoff,hep-ph/0203084.\\
8. M.Gluk, E.Reya  and A.Vogt, Z.Phys.C {\bf67}, 433(1995 ); Euro.Phys.J.C {\bf5}, 461(1998 ).\\
9.F.D. Aaron et al. [H1 Collaboration], phys.Lett.B\textbf{665},
139(2008); Eur.Phys.J.C\textbf{71},1579(2011). C.Adloff et al. [H1
Collaboration], Eur.Phys.J.C\textbf{21}, 33(2001).\\
10. S.Moch, J.A.M.Vermaseren, A.vogt, Phys.Lett.B \textbf{606},
123(2005); A. Vogt, S. Moch and J. A. M. Vermaseren, Nucl. Phys. B\textbf{691}, 129(2004).\\
11. A. D. Martin, R. G. Roberts, W. J. Stirling and R. Thorne,
Phys. Lett. B\textbf{531}, 216(2001).\\

\begin{figure}
\includegraphics[width=0.5\textwidth]{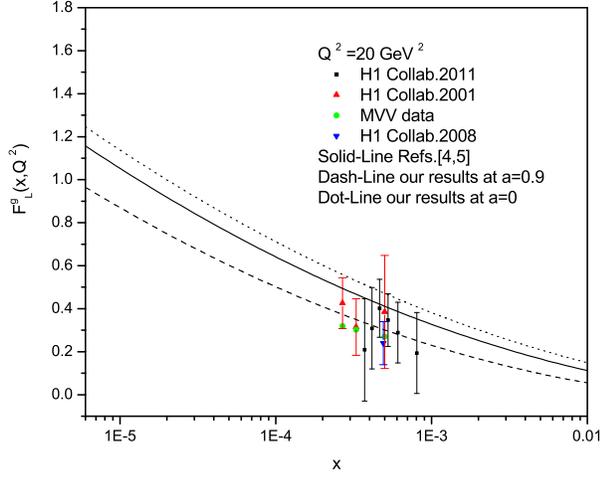}
\caption{LO longitudinal structure function at $Q^{2}=20 GeV^{2}$
according to the gluon distribution input [5] compared with
results Ref.[2], H1 data [9] and MVV [10] prediction. Our results
at the point of expansion $a=0$ shown with dot-line and  at the
point of expansion $a=0.9$ shown with dash-line.}\label{Fig1}
\end{figure}
\begin{figure}
\includegraphics[width=0.5\textwidth]{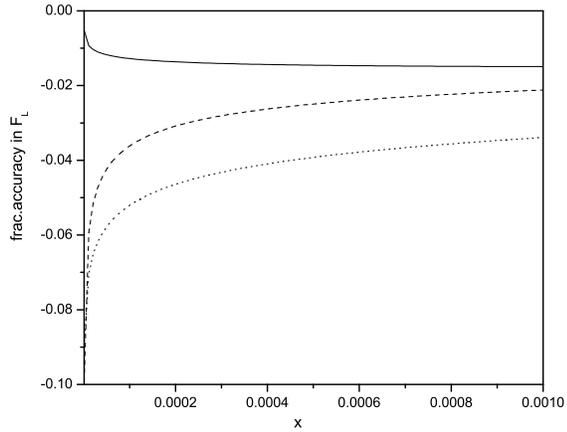}
\caption{Fractional accuracy plots for $F_{L}$ using input gluon
distributions [5-8] with respect to the NLO QCD parameterizations
of MRST [11] for $Q^{2}=20 GeV^{2}$. The solid ($F_{Block,MRST}$),
dash ($F_{DL,MRST}$) and dot ($F_{GRV,MRST}$) curves are taken
from the Block model, DL model and GRV parameterization
respectively. }\label{Fig2}
\end{figure}
\begin{figure}
\includegraphics[width=0.5\textwidth]{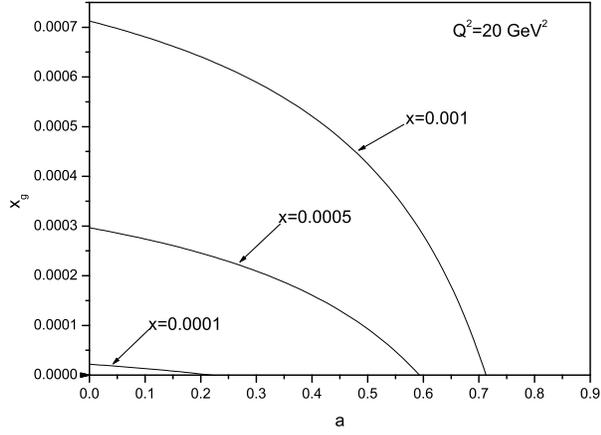}
\caption{Behaviour of $x_{g}$($=kx$,
$k=\frac{1}{1-a}(1-a+\frac{B}{A})$) vs the points of expansion $a$
at NLO.}\label{Fig3}
\end{figure}
\begin{figure}
\includegraphics[width=0.5\textwidth]{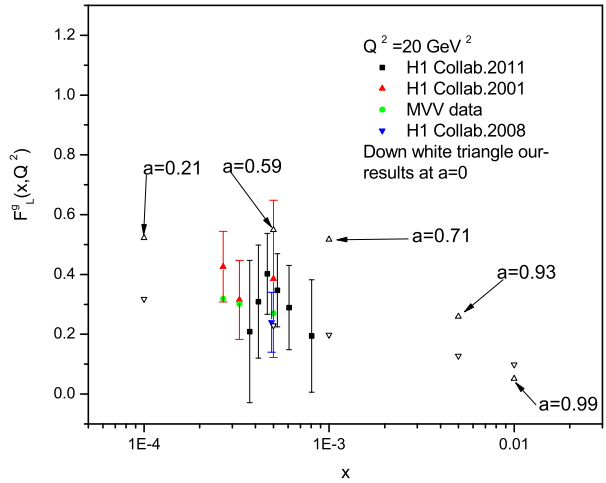}
\caption{NLO longitudinal structure function at $Q^{2}=20 GeV^{2}$
according to the gluon distribution input [5] compared with
results Ref.[2], H1 data [9] and MVV [10] prediction. Our results
at the point of expansion $a=0$ shown with down- white triangle
and at the  maximum point of  expansion shown with up- white
triangle.}\label{Fig4}
\end{figure}
\begin{figure}
\includegraphics[width=0.5\textwidth]{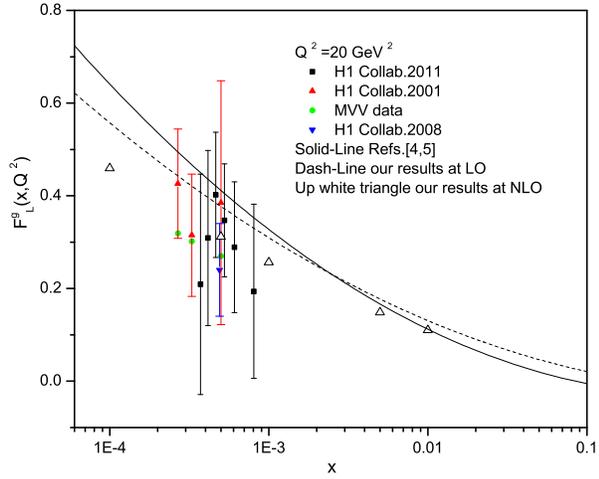}
\caption{Behaviour of the LO and NLO longitudinal structure
function at $Q^{2}=20 GeV^{2}$ according to the gluon distribution
input [5] compared with results Ref.[2](solid- line), H1 data [9]
and MVV [10] prediction. Our results according to the hard-Pomeron
behaviour at LO  shown with dash-line and  at NLO shown with up-
white triangle.}\label{Fig5}
\end{figure}


\end{document}